\documentclass[a4paper,12pt]{article}

\usepackage{color}
\usepackage[utf8]{inputenc}
\usepackage[english]{babel}
\usepackage{amssymb,amsmath}
\usepackage[pdftex]{graphicx}
\usepackage[pdftex]{hyperref}
\usepackage[left=3cm,right=2.5cm,top=2cm,bottom=3cm,bindingoffset=0cm]{geometry}

\begin{document}
\renewcommand\bibname{\Large References}

\title{
\begin{flushright}
{\small INR-TH-2015-006}
\end{flushright}
Towards the correspondence between Q-clouds and sphalerons}

\author{
E.\;Nugaev$^a$\thanks{{\bf e-mail}:
emin@ms2.inr.ac.ru}, A.\;Shkerin$^{a,b}$\thanks{{\bf e-mail}:
shkerin@inr.ru}
\\
$^a${\small{\em
Institute for Nuclear Research of the Russian Academy
of Sciences,}}\\
{\small{\em 60th October Anniversary prospect 7a, 117312, Moscow,
Russia
}}\\
$^b${\small{\em \'{E}cole Polytechnique F\'{e}d\'{e}rale de Lausanne,
}}\\
{\small{\em  CH-1015, Lausanne, Switzerland}}}
\date{}

\maketitle

\begin{abstract}
Non-linear classical equations of motion may admit degenerate solutions at fixed charges. While the solutions with lower energies are classically stable, the ones
with larger energies are unstable and refereed as Q-clouds. We consider a theory
in which the homogeneous charged condensate is classically stable and argue that Q-clouds correspond to sphalerons between the stable Q-balls and the condensate. For the model
with analytic solution, we present Arrhenius formula for the quantum production
of Q-balls from the condensate at large temperatures.
\end{abstract}

\section{Introduction}

In the theory of complex scalar field with global U(1)-invariance, there are localized stationary solutions of the classical equations of motion. These solutions
of non-linear equations correspond to nontopological solitons, see~\cite{Lee:1991ax} for review. In the theory of the single scalar field in three spatial dimensions, they are also refereed as Q-balls and thoroughly investigated in~\cite{Coleman:1985ki}. The numerous applications of Q-balls in cosmology (see, for example, the book~\cite{gorbunov2011introduction}) rise a problem of their production, which can be solved in different ways~\cite{Kusenko:1997hj,Kasuya:2000wx,Khlebnikov:1999qy,Krylov:2013qe,Amin:2010xe}. The crucial feature of these mechanisms is the presence of the classically unstable homogeneous charged condensate, whose decay leads to formation of (quasi)stable localized configurations. The classical clumping of the condensate is, thus, the dominant channel of nontopological solitons production.

In this paper we revise the conditions of classical stability of the solutions in theories with scalar field potentials. We show that certain class of potentials admits the classically stable condensate as well as stable Q-ball configurations. This implies the possible existence of another mechanism of transitions between these solutions~\cite{Lee:1985uv}. Typically, there are two localized solutions for the same charge in the wide range of the latter. One of them may correspond to stable Q-ball, another can represent an unstable configuration, which was refereed as Q-cloud~\cite{Alford:1987vs}. Stable and unstable solutions are arranged into branches in energy-charge coordinates. In this case, the upper branch contains Q-clouds, and the lower branch consists of Q-balls. In the finite-volume theory there are also branches of the condensate solutions which are also degenerate in charge. We will argue that Q-clouds correspond to the sphalerons between two stable solutions with equal charge. Q-balls production from the stable condensate is, therefore, possible at least at finite temperatures. The rate of production is determined by Arrhenius formula, and the sphaleron energy is a crucial quantity in its evaluation.

%In this case there are two branches of localized 
%solutions. The branch with the lower energies are classically stable Q-balls and
%the unstable upper branch, which were refereed as Q-clouds in~\cite{Alford:%1987vs}. We will argue that the Q-balls production from the stable condensate is %possible for the finite temperatures. The Q-clouds correspond to the sphalerons %between two stable solutions for different charges. The rate of quantum %production is determined by Arrhenius formula. The crucial value for the later ---
%sphaleron energy is calculated in the simple model.

The paper is organized as follows: In Section 2 we summarize features of Q-balls
using the dependence of soliton's energy on its charge. In Section 3 the condensate
stability issue is revisited. In Section 4 we will examine small excitations around nontopological solitons in the toy model. In the Section 5 we present Arrhenius formula for the theories in one and three spatial dimensions.

\section{Different branches for classical solutions}

In this section we will consider some properties of Q-balls in infinite and
flat four-dimensional space-time. We will use the following form of Lagrangian
for the theory of complex scalar field $\phi$, 
\begin{equation}
{\cal L}=\partial_\mu\phi^*\partial^\mu\phi-V(|\phi|).
\label{L}
\end{equation}
The potential $V$ is assumed to be function of $U(1)$-invariant
combination $\phi^*\phi$, however, for the convenience we define it here through the explicit dependence on the field moduli. The general conditions for the
existence of localized stationary configurations of the form
\begin{equation}
\phi(t,\vec{x})=f(r){\it e}^{{\it i}\omega t}
\label{ansatz}
\end{equation}
were derived in \cite{Coleman:1985ki}. As was shown in \cite{Rosen}, which is  probably the first work
on Q-balls, for the parabolic-piecewise potential
%\footnote{We suppose
%$m^2\ge 0$,  otherwise configurations with large charge have different %properties, see}
\begin{equation}
V(|\phi|)=M^2\phi^*\phi\,\theta\left(1-\frac{\phi^*\phi}{v^2}\right)+(m^2\phi^*\phi+v^2(M^2-m^2))\theta\left(\frac{\phi^*\phi}{v^2}-1\right)
\label{potential}
\end{equation} 
one can obtain exact solution. Moreover, there is a possibility for the explicit
analysis of excitations around this solution \cite{Gulamov:2013ema}.
  
\begin{figure}[!h]
\center{\includegraphics[width=0.6\linewidth]{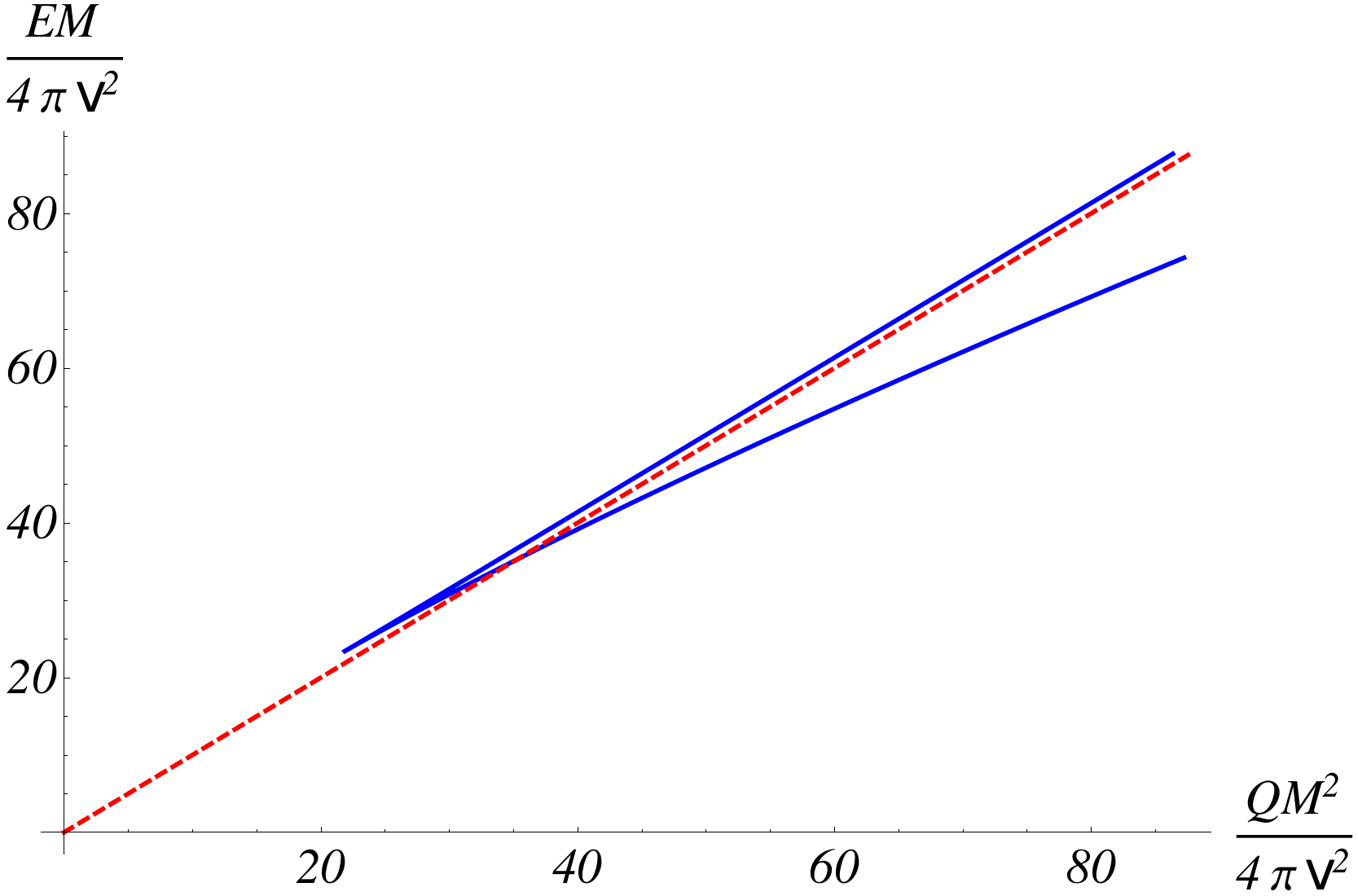}}
\caption{$E(Q)$ plot for the
Q-ball in (3+1)-dimensional theory, $m^2>0$, $M/m=10$. Dashed line
corresponds to free particles with $E=MQ$.} \label{3D}
\end{figure}

To describe general properties of Q-balls, it is useful to consider the  dependence of the energy $E$ of the soliton on its charge $Q$. For  ansatz~(\ref{ansatz}), they are given by
\begin{equation}\label{energy and charge}
E=\int d^3x(\omega^2f^2+(\partial_i f)^2+V(f)),\;\;\; Q=2\omega\int d^3xf^2.
\end{equation} 
We are interested in non-negative values of $m^2$, $m^2\ge 0$. However, it should be mentioned that in the opposite case, $m^2<0$, the
properties of the stable solutions, if they exist, are different\footnote{Negative values of $m^2$ can be used for qualitative analysis of Q-balls in the theory with an additional vacuum,  $\phi^*\phi\neq 0$, of potential $V$, see \cite{Nugaev:2013poa}} only for large $Q$. The plot shown in Fig.[\ref{3D}] is a very typical for the wide class of nontopological solitons, see 
\cite{PhysRevD.13.2739} for example. There is upper branch of unstable Q-clouds 
\cite{Alford:1987vs}, and the branch of stable solutions with lesser energies for the same charge. 

The less trivial characteristic of Q-clouds is the number of unstable modes. In the work \cite{Gulamov:2013ema} only one mode was found for the potential (\ref{potential}).
 
\section{Stability of the condensate}

Classical equations of motion also have spatially homogeneous solutions. However, to obtain the finite values of their charge and energy, one should regularize the theory by putting it on a compact manifold. We will consider the theory in one spatial dimension and assume the periodic boundary conditions for $\phi$ and $\phi^*$. 
Let us study the charged solutions of the form
\begin{equation}
\phi_0=C{\it e}^{{\it i}\Omega t},
\label{condensate}
\end{equation}
where $C$ is some constant, and $\Omega$ is determined from the equations of motion, 
\[
\Omega^2=\dfrac{1}{2C}V'(C).
\]
The charge and the energy of the configuration (\ref{condensate}) are expressed through the constant $C$ and the size $L$ of spatial $S^1$ manifold. Crucial feature of charged condensate for the classical production of solitons was caught in \cite{Zakharov} for the model with negative quartic coupling. In this case, classical instability of (\ref{condensate}) corresponds to the instability of the condensate in a Bose gas with attraction. Analysis for arbitrary potential $V$, see, for example \cite{gorbunov2011introduction},
gives the following condition of instability of (\ref{condensate}),\footnote{Generalization of this condition for the expanding Universe is also possible \cite{Kasuya:2000wx}.}
\begin{equation}
V''(C)-\frac{V'(C)}{C}<0
\label{condition}
\end{equation}
For the potential (\ref{potential}) this condition is obviously violated, and the condensate is stable. However, the choice of parabolic-piecewise potential is not crucial for the classical stability of (\ref{condensate}). For example, one can regularize $\theta$-functions in (\ref{potential}), according to the rule
\begin{equation}
\theta\left(\frac{\phi^*\phi}{v^2}-1\right)\to\frac{1}{2}\left(1+\tanh{\left[
\alpha\left(\frac{\phi^*\phi}{v^2}-1\right)\right]}\right),
\label{regularization}
\end{equation}
where $\alpha$ is some large constant. The condition (\ref{condition}) is still violated for $\vert C\vert<v$, and the charge of such stable condensate can be arbitrary large in the limit $L\to\infty$. Thus, the classical
production of Q-balls from these condensate solutions is forbidden, although the solitons are energetically more preferable.

It should be noted that violation of the condition (\ref{condition}) for small
amplitudes of the field for the potential (\ref{potential}) does not mean the 
absence of Q-ball solution. Thus, for large charges there is a possibility
of the simultaneous stability of homogeneous charged condensate and spatially localized soliton.

\section{A toy model: nontopological solitons on a circle}

We now turn to the simple model in which we will illustrate the relations between stable Q-balls, unstable Q-clouds, and stable condensate. For this, we reduce the number of spatial dimensions to one, which is parametrized by$x$. The ansatz reads as follows
\begin{equation}
\phi(x,t)=f(x){\it e}^{{\it i}\omega t}.
\label{ansatzS}
\end{equation}
The energy and the charge of this configuration are
\begin{equation}\label{energy and charge on S}
E=\int dx(\omega^2f^2+f'^2+V(f)),\;\;\; Q=2\omega\int dxf^2.
\end{equation}
After the substitution of (\ref{ansatzS}) into the classical equation of motion, one obtains,
\begin{equation}\label{Equation of Motion for f}
f''+\omega^2f=\dfrac{dV}{d(f^2)}f.
\end{equation}
We assume that the length of the circle is $L$ and $x\in(-L/2,L/2)$. With the potential (\ref{potential}), the periodic solution $f(-L/2)=f(L/2)$ is given by
\begin{equation}\label{Q-ball}
f=\left\lbrace\begin{array}{l}
v\dfrac{\cos ax}{\cos ax_0},\;\;\vert f\vert>v,\;\;a^2=\omega^2-m^2>0\\
v\dfrac{\cosh bx}{\cosh bx_0},\;\;\vert f\vert<v,\;\;b^2=M^2-\omega^2>0,
\end{array}\right.
\end{equation}
where $x_0$ is a matching point, defined by equation
\begin{equation*}
a\tanh a(x_0-L/2)=-b\tan bx_0,
\end{equation*}
which is obtained by requirement of continuity of $f$ and its derivative at this point.
The shape of the function $f$ depends on the particular choice of $\omega$, for which we require $m<\omega<M$. So, one can formally write $f=f(x,\omega)$. Using the expressions (\ref{energy and charge on S}) one can obtain
\begin{equation}\label{dE/dQ}
\dfrac{dE}{d\omega}=\omega\dfrac{dQ}{d\omega}.
\end{equation}
This condition is known to hold for many various Q-ball solutions (\cite{tsumagari2009physics},\cite{PhysRevD.13.2739},\cite{Lee:1991ax}). It provides a useful check for numerical calculations, and indicates that $E$ and $Q$, as functions of $\omega$, can have only simultaneous extrema.

\begin{figure}[h!]
\begin{minipage}[h]{0.49\linewidth}
\center{\includegraphics[width=0.99\linewidth]{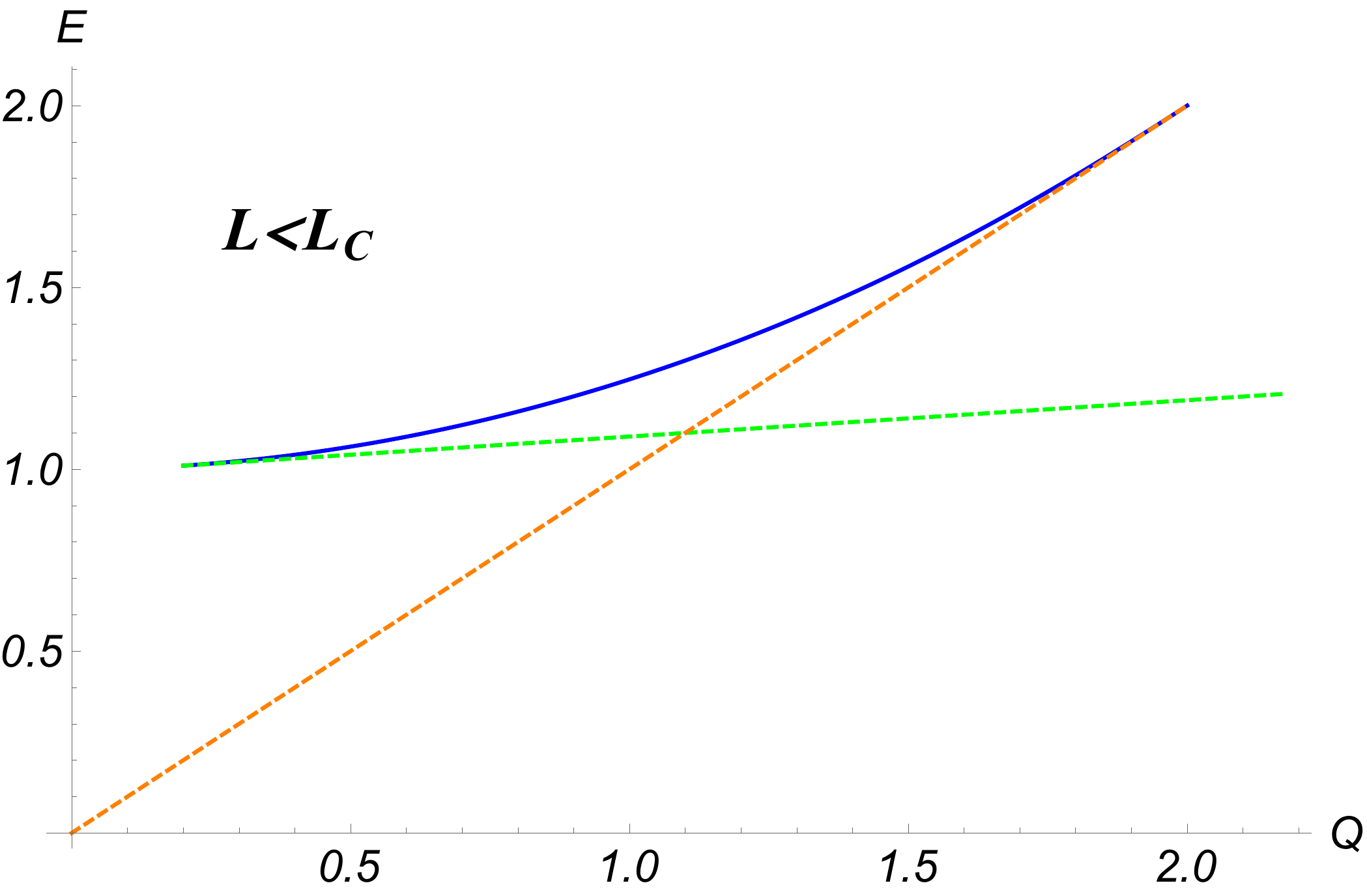}\\a}
\end{minipage}
\hfill
\begin{minipage}[h]{0.49\linewidth}
\center{\includegraphics[width=0.99\linewidth]{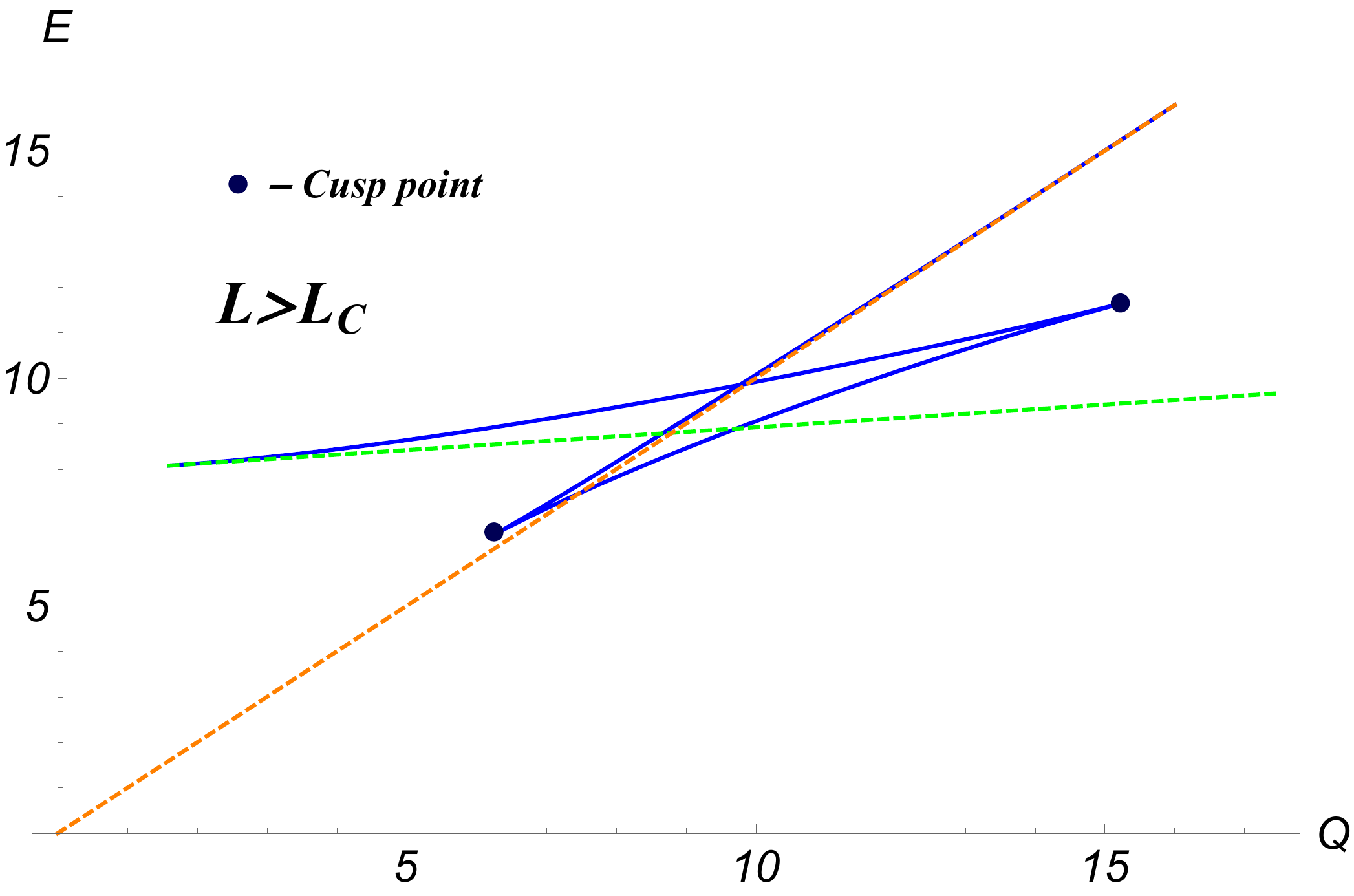}\\b}
\end{minipage}
\caption{ $E(Q)$ plots for the different values of $L$. The dashed lines represent the condensate solutions, the solid lines - Q-balls. We put $M=v=1$, and $m=0.1$. a) $L=1$, b) $L=8$.}
\label{E(Q)-plots}
\end{figure}

In the case we are interested in, $m^2>0$, there are two families of stable condensate solutions:
\begin{equation}\label{Condensate}
\phi=\left\lbrace\begin{array}{l}
C{\it e}^{{\it i}Mt},\;\;\vert C\vert<v,\\
C{\it e}^{{\it i}mt},\;\;\vert C\vert>v.
\end{array}\right.
\end{equation}
The important relations between them and Q-balls are established by investigation of their $E(Q)$ dependence. On the Fig.(\ref{E(Q)-plots}), we present this dependence for the different choice of parameters. The properties of Q-balls depend significantly on the value of $L$. For $L<L_c\sim 1/M$, they form one branch, whose ends are attached to the condensate lines. For $L>L_c$, the picture is qualitatively different. There are now three branches of Q-balls with different signs of ${\it d}^2E/{\it d}Q^2$. As we will see, the one with the negative sign is stable. The cusp points, separating the branches, correspond to the simultaneous extrema of ${\it d}Q/{\it d}\omega$ and ${\it d}E/{\it d}\omega$, in agreement with Eq.(\ref{dE/dQ}). 

In the limit $L\rightarrow\infty$ the only relevant solutions are those which keep the finite values of $E$ and $Q$. It can be shown that only the vicinity of the lower cusp is relevant in this case. Indeed, for the homogeneous solutions (\ref{Condensate}), both $E$ and $Q$ are proportional to the size $L$, in contrast with the lower branch of Q-balls. Certainly, the dependence $E(Q)$ for relatively small charges in this limit reproduces Fig.(\ref{3D}) for infinite space\footnote{For the potential (\ref{potential}), the dependences $E(Q)$ are qualitatively similar in the case of one and three spatial dimensions \cite{Gulamov:2013ema}.}.

Our matter of interest is the stability issue of the relevant solutions with equal charge, in the limit $L\rightarrow\infty$. For this purpose, the potential (\ref{potential}) is very useful, since it allows the analytical investigation not only for the background solutions, but also for small excitations
around them.

Let us justify the notion of stable and unstable branches used before. 
We will use the following ansatz \cite{Anderson:1970et} for excitations
\begin{equation}\label{Perturbation ansatz}
h(x,t)=e^{i\Omega t}\left(f_1(x)e^{i\alpha t}+f_2^*(x)e^{-i\alpha^*t}\right),
\end{equation}
where $f_{1,2}$ are complex functions, $\alpha=-i\gamma+\gamma'$; $\gamma,\gamma'\in\texttt{R}$.
For the background Q-ball solution (\ref{ansatzS}),
the linearised equations for perturbations are
\begin{equation}\label{Linearised equations}
\left\lbrace
\begin{array}{l}
f_1''+(\omega+\alpha)^2f_1=Uf_1+(m^2-M^2)\delta\left(1-\dfrac{f^2}{v^2}\right)(f_1+f_2),\\
f_2''+(\omega-\alpha)^2f_2=Uf_2+(m^2-M^2)\delta\left(1-\dfrac{f^2}{v^2}\right)(f_1+f_2),
\end{array}\right.
\end{equation} 
\begin{equation*}
U=M^2\theta(x_0-x)+m^2\theta(x-x_0).
\end{equation*}
The analysis of this system can be reduced to the consideration of a system
of linear equations. %see \cite{Gulamov:2013ema,Nugaev:2014iva}.
We find solutions of Eq.(\ref{Linearised equations}) for $|x|<x_0$ and $x_0<|x|<L/2$ separately, and match them at the point $x=x_0$. Due to the presence of delta-functions, these solutions are continuous but not smooth at this point. In each interval, the general solution of Eq.(\ref{Linearised equations}) contains two arbitrary constants. The boundary conditions fix one of them. At $x=x_0$, Eq.(\ref{Linearised equations}) form a system of linear homogeneous equations on the rest of the constants. Hence, the solution exists, if the determinant of this system $\Delta=\Delta(\alpha)$ equals zero for some value of $\alpha$.

Equation $\Delta(\alpha)=0$ has at least one solution, $\alpha=0$, which corresponds to zero modes. One of them has the form $h\sim i\Phi_0$ and arises due to U(1) symmetry of the action. One more mode appears as a result of breaking of the translational invariance by the Q-ball configuration, $h\sim\Phi'_0$. This implies $\Delta(0)=\Delta'(0)=0$. Calculations show that the existence of nonzero root of $\Delta(\alpha)$ is correlated with the sign of $\partial^2E/\partial Q^2$. Namely, for Q-clouds with $\partial^2E/\partial Q^2>0$ there is exactly one purely imaginary root $\alpha=-i\gamma$. For solitons with $\partial^2E/\partial Q^2<0$ no roots were found. Thus, we can indeed speak about stable and unstable branches of $E(Q)$ plots. In the limit $L\rightarrow\infty$, this matches with the results of the work \cite{Gulamov:2013ema}. 

\begin{figure}[h!]
\begin{minipage}[h]{0.49\linewidth}
\center{\includegraphics[width=0.99\linewidth]{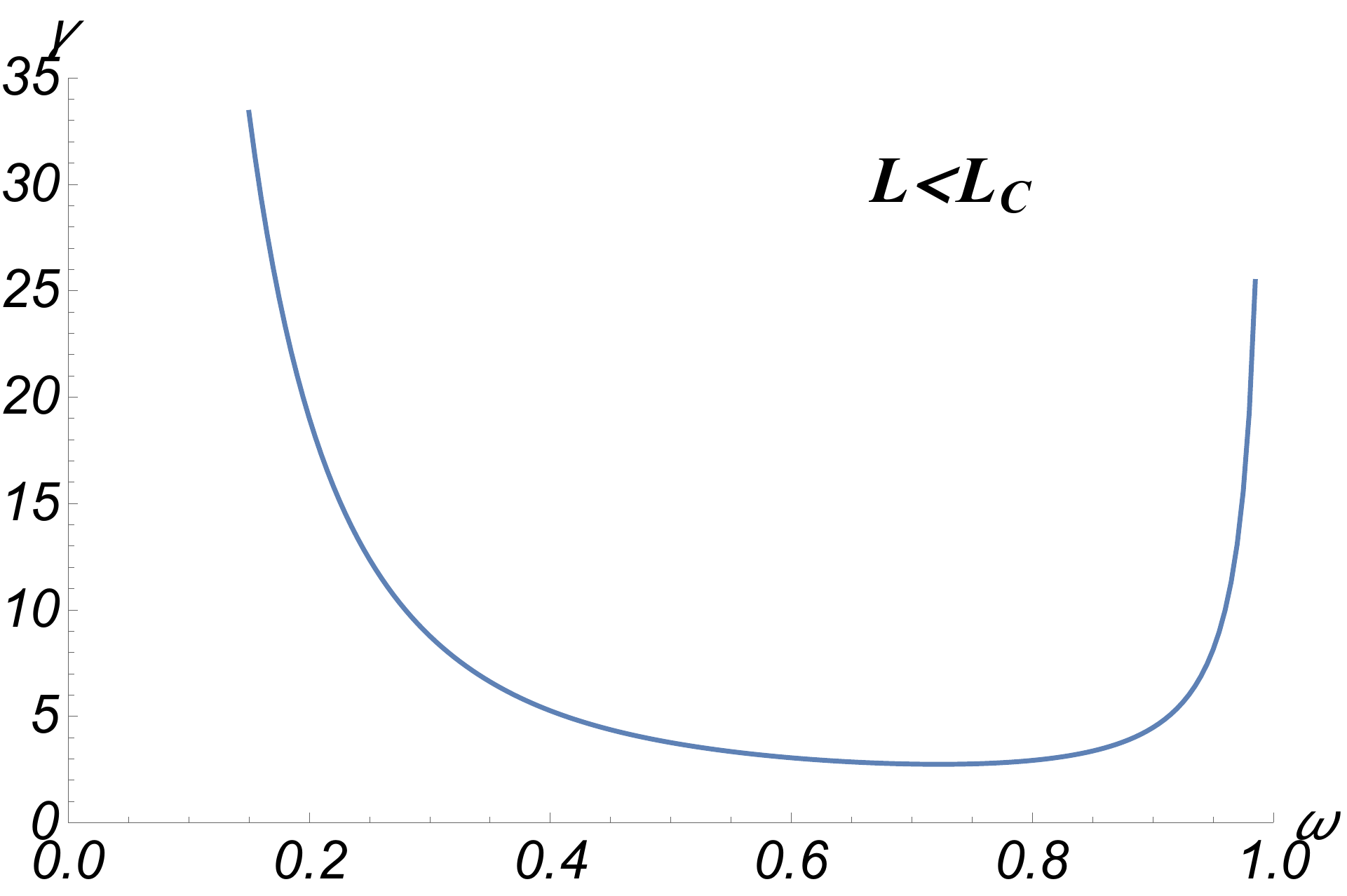}\\a}
\end{minipage}
\hfill
\begin{minipage}[h]{0.49\linewidth}
\center{\includegraphics[width=0.99\linewidth]{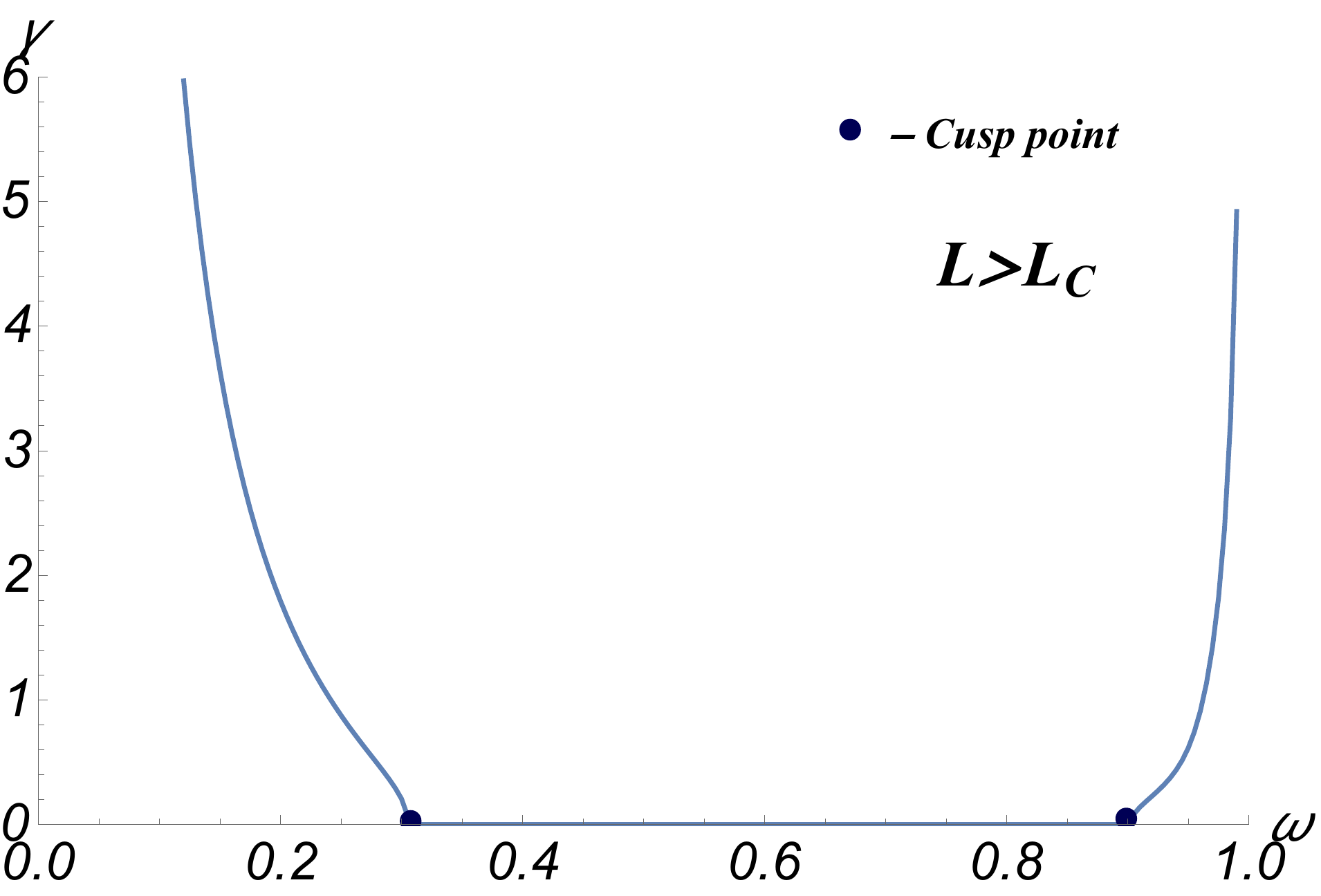}\\b}
\end{minipage}
\caption{Decay rate of Q-balls for the different values of $L$. We assume $M=v=1$, and $m=0$. a) $L=1$, b) $L=8$.}
\label{Negative root plots}
\end{figure}

The dependence of $\gamma$ for the particular Q-ball solution is shown in Fig.(\ref{Negative root plots}). As $\omega$ approaches $m$ or $M$, the life time $\tau\sim\gamma^{-1}$ of the solution tends to zero. This strengthens our suspicion of the existence of the classical path connecting such a Q-ball and the condensate solution lying directly below it.

\section{Sphaleron transitions}

\begin{figure}[h!]
\begin{minipage}[h]{0.49\linewidth}
\center{\includegraphics[width=0.99\linewidth]{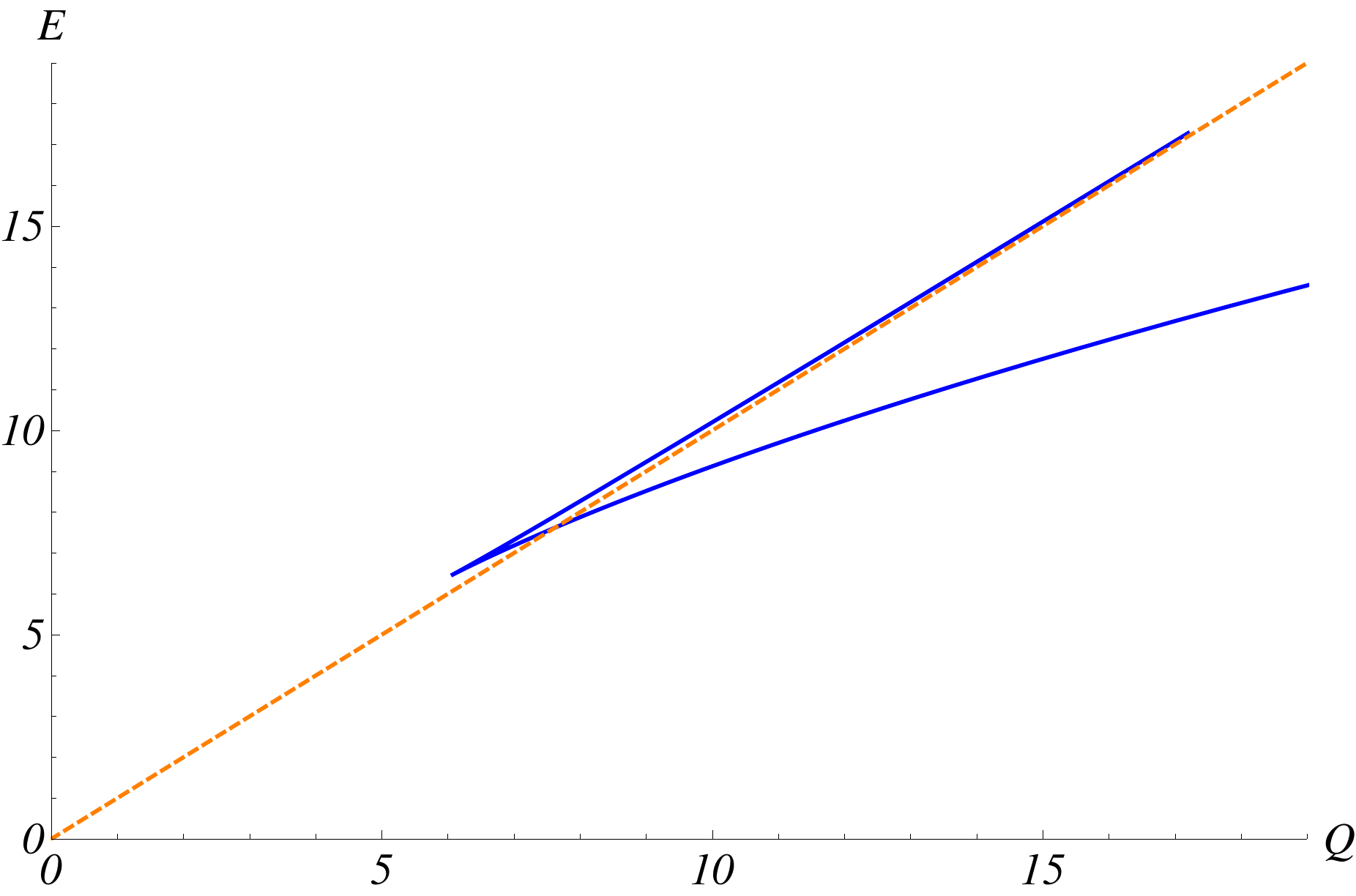}\\a}
\end{minipage}
\hfill
\begin{minipage}[h]{0.49\linewidth}
\center{\includegraphics[width=0.99\linewidth]{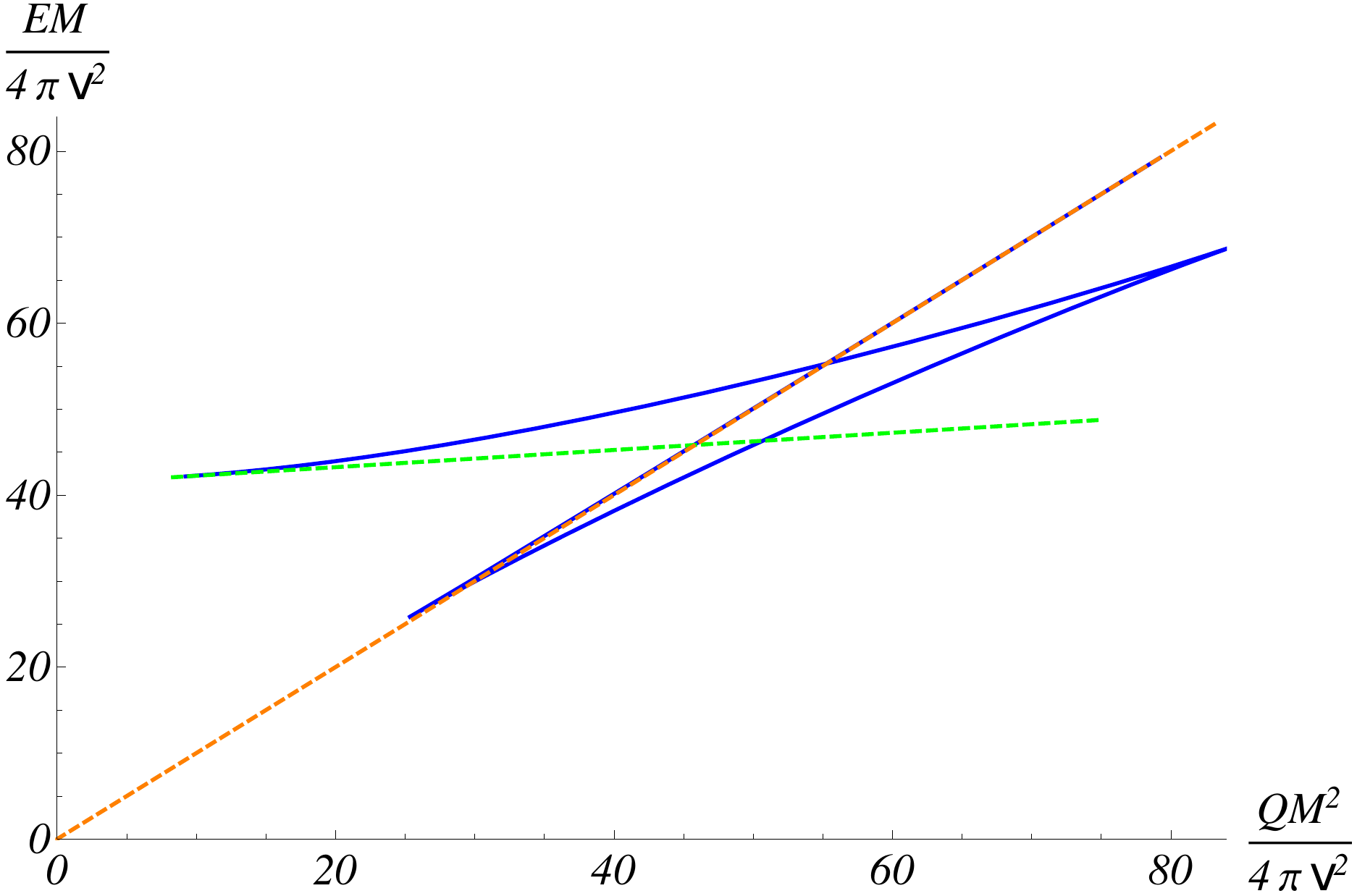}\\b}
\end{minipage}
\caption{ a) $E(Q)$ plot for the (1+1)-dimensional compactified Q-balls with finite values of $Q$ and $E$, in the limit of large $L$. The parameters of the example: $M=v=1$, $m=0.1$, $L=20$. 
b) $E(Q)$ plot for the
Q-balls living on the 3-dimensional sphere of circumference $L=10$, $m^2>0$, $M/m=10$. Dashed lines correspond to the condensate solutions.}
\label{3D-1}
\end{figure}

If $Q$ is large enough, then there are three
solutions of the same charge in the compactified model in the limit $L\rightarrow\infty$, as shown on the Fig.(\ref{3D-1}a). Two of
them are stable, and one corresponds to unstable Q-cloud.
The presence of precisely one negative mode for Q-cloud allows their interpretation as sphalerons. They are saddle points between
two stable solutions. The sphalerons play key role in transitions between
stable configurations at large temperatures \cite{Rubakov:1996vz}.
It should be mentioned that it is the compactification of the theory that  
makes our analysis appropriate, since for the infinite space, $E$ and $Q$ of the condensate are unbounded below. Setting the model on the circle also clarifies sphaleron physics for time-independent vacua \cite{Manton:1988az}. 

%From Fig.(\ref{E(Q)-plots}) is is clear that for $\vert C\vert$ sufficiently %close to $v$, the height of the barrier between the condensate and the Q-cloud %can be made as small as desired. In the large $L$ limit this corresponds to %choosing the solutions with sufficiently large $Q$, or, equivalently, with %$b\equiv\sqrt{M^2-\omega^2}\ll\sqrt{M^2-m^2}$. This is not a particular feature %of (1+1)-dimensional Q-balls. In a realistic (3+1)-dimensional case the energy %gap between these two solutions is also arbitrary small as far as $Q$ is large %enough (\cite{Gulamov:2013ema}). 

Now, let us put our system into a thermal bath with the temperature $T$. Choose the charge $Q_0$ for which (and for any $Q>Q_0$) $T\gg\Delta E=E_{Q-cloud}-E_{condensate}$. In this case, the dominant transition channel from the condensate to Q-ball occurs via sphaleron. To the leading order, the decay rate of the condensate is given by Arrhenius formula, $\Gamma\sim\exp(-\Delta E/T)$. In the limit of small $b$ it reads,
\begin{equation}\label{Gamma 1+1}
\Gamma_{1+1}\sim e^{-\dfrac{v^2 b}{2T}}\left(1+O\left(\frac{b^2}{T}\right)\right).
\end{equation}
So far we have considered (1+1)-dimensional periodic solutions. It is not difficult, however, to perform the same calculation for (3+1)-dimensional periodic solutions, see Fig.(\ref{3D-1}b). Under the same assumptions, this leads to the result,
\begin{equation}\label{Gamma 3+1}
\Gamma_{3+1}\sim e^{-\dfrac{\pi^3 v^2 b}{2(M^2-m^2)T}}\left(1+O\left(\frac{b^2}{T}\right)\right).
\end{equation}
Let us clarify the applicability conditions of the formulas (\ref{Gamma 1+1}) and (\ref{Gamma 3+1}). They describe the transition from one stable configuration to another, that is, the theory should allow the existence of two stable and one unstable solutions at certain values of $Q$. The charge conservation determines 
the field amplitude $C$ of initial configuration through $Q=2MC^2L^3$.
Secondly, the temperature should be high compared with the height of the barrier. Finally, there should be thermal equilibrium among other species coexisting with the field $\phi$.

\section{Conclusion and Acknowledgements}

In the infinite space-time, transitions from the classically stable condensate of a charge $Q=2MC^2L^3$ to the stable
Q-ball of the same charge at finite temperature can be described using the unstable Q-cloud solution.
For the large enough temperatures the rate of transitions (\ref{Gamma 3+1}) in the unit time in the volume $L^3$ can be used. We clarify our consideration by the compactification of the theory where both initial and final configurations can be considered as solutions of the classical equations of motion with the finite energies.

We would like to thank V.~A.~Rubakov, M.~E.~Shaposhnikov, M.~N.~Smolyakov and I.~I.~Tkachev for
helpful discussions. This work is supported by Russian Science Foundation grant
14-22-00161.

\bibliography{Periodic_Q_balls}

\begin{thebibliography}{10}

\bibitem{Lee:1991ax}
T.D. Lee and Y.~Pang.
\newblock {Nontopological solitons}.
\newblock {\em Phys.Rept.}, 221:251--350, 1992.

\bibitem{Coleman:1985ki}
Sidney~R. Coleman.
\newblock {Q Balls}.
\newblock {\em Nucl.Phys.}, B262:263, 1985.

\bibitem{gorbunov2011introduction}
D.S. Gorbunov and V.A. Rubakov.
\newblock {\em Introduction to the Theory of the Early Universe: Hot Big Bang
  Theory}.
\newblock World Scientific, 2011.

\bibitem{Kusenko:1997hj}
Alexander Kusenko.
\newblock {Phase transitions precipitated by solitosynthesis}.
\newblock {\em Phys.Lett.}, B406:26--33, 1997.

\bibitem{Kasuya:2000wx}
S.~Kasuya and M.~Kawasaki.
\newblock {Q Ball formation in the gravity mediated SUSY breaking scenario}.
\newblock {\em Phys.Rev.}, D62:023512, 2000.

\bibitem{Khlebnikov:1999qy}
S.~Khlebnikov and I.~Tkachev.
\newblock {Quantum dew}.
\newblock {\em Phys.Rev.}, D61:083517, 2000.

\bibitem{Krylov:2013qe}
E.~Krylov, A.~Levin, and V.~Rubakov.
\newblock {Cosmological phase transition, baryon asymmetry and dark matter
  Q-balls}.
\newblock {\em Phys.Rev.}, D87(8):083528, 2013.

\bibitem{Amin:2010xe}
Mustafa~A. Amin.
\newblock {Inflaton fragmentation: Emergence of pseudo-stable inflaton lumps
  (oscillons) after inflation}.
\newblock 2010.

\bibitem{Lee:1985uv}
Ki-Myeong Lee and Erick~J. Weinberg.
\newblock {Tunneling without barriers}.
\newblock {\em Nucl.Phys.}, B267:181, 1986.

\bibitem{Alford:1987vs}
Mark~G. Alford.
\newblock {Q clouds}.
\newblock {\em Nucl.Phys.}, B298:323, 1988.

\bibitem{Rosen}
G.~Rosen.
\newblock {Particlelike Solutions to Nonlinear Complex Scalar Field Theories
  with Positive-Definite Energy Densities}.
\newblock {\em J.Math.Phys.}, 9:996--999, 1968.

\bibitem{Gulamov:2013ema}
I.E. Gulamov, E.~Ya. Nugaev, and M.N. Smolyakov.
\newblock {Analytic $Q$-ball solutions and their stability in a piecewise
  parabolic potential}.
\newblock {\em Phys.Rev.}, D87(8):085043, 2013.

\bibitem{Nugaev:2013poa}
E.~Ya. Nugaev and M.N. Smolyakov.
\newblock {Particle-like Q-balls}.
\newblock {\em JHEP}, 1407:009, 2014.

\bibitem{PhysRevD.13.2739}
R.~Friedberg, T.~D. Lee, and A.~Sirlin.
\newblock Class of scalar-field soliton solutions in three space dimensions.
\newblock {\em Phys. Rev. D}, 13:2739--2761, May 1976.

\bibitem{Zakharov}
V.~E. Zakharov.
\newblock {\em JETP}, 26:994, 1968.

\bibitem{tsumagari2009physics}
Mitsuo~I Tsumagari.
\newblock {The Physics of Q-balls}.
\newblock {\em arXiv preprint arXiv:0910.3845}, 2009.

\bibitem{Anderson:1970et}
D.L.T. Anderson and G.H. Derrick.
\newblock {Stability of time-dependent particlelike solutions in nonlinear
  field theories. 1.}
\newblock {\em J.Math.Phys.}, 11:1336--1346, 1970.

\bibitem{Rubakov:1996vz}
V.A. Rubakov and M.E. Shaposhnikov.
\newblock {Electroweak baryon number nonconservation in the early universe and
  in high-energy collisions}.
\newblock {\em Usp.Fiz.Nauk}, 166:493--537, 1996.

\bibitem{Manton:1988az}
N.S. Manton and T.M. Samols.
\newblock {Sphalerons on a circle}.
\newblock {\em Phys.Lett.}, B207:179, 1988.

\end{thebibliography}
\bibliographystyle{unsrt}

\end{document}